\documentclass[twocolumn,aps,prl]{revtex4-1}
\usepackage{graphicx}
\usepackage{ulem}
\begin{document}

\title{Pauli Blocking Effect on Efimov States Near Feshbach Resonance}
\author{David James MacNeill}
\altaffiliation[Current address: ]{Department of Physics, Cornell University, Ithaca, NY 14853, USA}
\author{Fei Zhou}
\affiliation{Department of Physics and Astronomy, The University of British Columbia, Vancouver, B. C., Canada V6T1Z1}

\date{{\small Sept 10, 2010}}

\begin{abstract}
In this Letter we study the effect of Pauli blocking on Efimov states in
a quantum Fermi gas and illustrate that
the universal Efimov potential is altered at large distances.
We obtain the universal spectrum flow of Efimov trimers
when the Fermi density is varied and further consider the effect of
scattering of trimers by the Fermi sea. We
argue that the universal flow is robust against
fluctuating particle-hole pairs
that result in an infrared catastrophe
in impurity problems.
\end{abstract}

\maketitle

Recently, universal three-body Efimov structures that were first proposed in the 1970s have been successfully probed and determined in a series of remarkable cold-atom
experiments on inelastic loss spectra
\cite{Efimov70,Efimov72,Esry99,Braaten01,Kraemer06,Braaten06,Hammer07,Ottenstein08,Stecher09,Ferlaino09,Williams09,
Zaccanti09,Pollack09}.
The current data that cover an impressive range
of scattering lengths quite conclusively demonstrate the universality of the Efimov structures
\cite{Zaccanti09,Pollack09}.
The spectacular progress made on this subject has further illustrated
that the physics of cold atoms at sub-microkelvin temperatures can shed light on fundamental issues
in other systems at very different energy scales, in this case the few-body structures of
nuclei.

Although Efimov physics \cite{Efimov70,Efimov72}
and the theory of loss rates \cite{Esry99,Braaten01}
have so far been quite successful in explaining many aspects of the data,
in experiments on cold gases there are always many-body backgrounds.
Logically, it is important to understand how few-body states
respond to the presence of a quantum many-body background below
degeneracy temperatures.
For instance, can Efimov trimers survive a many-body background and
how are they affected when scattered by the background?
An equally important and challenging question is what kind of many-body correlations
can be induced by the universal few-body states
in a quantum gas or mixture.
In this Letter, we attempt to answer one of these questions,
and particularly to examine the
effect of Pauli blocking of scattering or open-channel Fermi atoms on the
three-body Efimov states.
To answer this question, we solve the three-body spectrum in the presence
of a Fermi sea and focus on the effect of Pauli blocking on
the Efimov physics.
These spectra are known to be one of the cornerstones for the theory of three-body
recombination and the
loss spectrum of metastable quantum gases. Our results can
also be applied to study the dynamics during the initial stage of
recombination and estimate the lifetime of quantum gases.

In the following, we consider that
two identical heavy Bose atoms with mass $M$ have
an interspecies resonance with a light Fermi atom with mass $m(\ll M)$
in the presence of a Fermi sea of light atoms. In the Born-Oppenheimer approximation, we can analyze
the fast motion of the light atom assuming that the two slow heavy atoms are a distance $R$ apart.
In the absence of a Fermi sea,
simultaneous near-resonance scattering of the light atom by the two heavy atoms induces a
bound state, and the binding energy at resonance is equal to $\hbar^2 \Omega^2 /2 \mu R^2$,
where $\Omega=0.567$ is the root of $\Omega=e^{-\Omega}$ (see below)
and $\mu=m M/(m+M)$ is the reduced mass of a light-heavy subsystem.
The bound state therefore glues together the two heavy atoms and yields an attractive long range potential $-\hbar^2\Omega^2/2 \mu R^2$.
This universal $1/R^2$ potential plays a paramount role in Efimov physics and results in the spectacular universal hierarchy structures in
three-body spectra \cite{Efimov70,Efimov72}.
We then investigate the effective attractive potential
between two heavy atoms in the presence of a Fermi sea
with Fermi momentum $\hbar k_{\text{F}}$.
The interspecies interactions can be treated as zero-range ones
\cite{Skorniakov57,Faddeev61}, so that for a light atom interacting with two heavy atoms at $\pm {\bf R}/2$, $V({\bf r})=V_0 [\delta({\bf r}-{\bf R}/2)+\delta({\bf r} +{\bf R}/2)]$ where

\begin{eqnarray}
\frac{1}{V_0}&=&\frac{\mu}{2\pi a_{\text{HL}}\hbar^2}-\frac{1}{(2\pi)^3}\int \! d\mathbf{k}\ \frac{1}{\epsilon^{\mu}_{k}}.
\label{scattering}
\end{eqnarray}
Eq.\ref{scattering} relates the contact interaction strength $V_0$ to
the scattering length $a_{\text{HL}}$ via a standard regularization procedure \cite{Cui10}. Here $\epsilon^\mu_k=\hbar^2 k^2/2\mu$ and
now the reduced mass $\mu \sim m$ since $M$ is much heavier than $m$.
The Schr{\"o}dinger equation for $\phi(\mathbf{k})$, the momentum-space
wavefunction for the light atom, is

\begin{eqnarray}
\left(\epsilon^{m}_{k}-E_{\text{L}}\right)\phi(\mathbf{k})=\frac{-2V_0}{(2\pi)^3}\int \! d\mathbf{k'} \cos\frac{(\mathbf{k}-\mathbf{k}')\cdot\mathbf{R}}{2} \phi(\mathbf{k}')
\end{eqnarray}
where $\epsilon^{m}_{k}=\frac{\hbar^2k^2}{2m}$ is the kinetic energy for
the light atom, and
$\phi(\mathbf{k})=0$ when $\mathbf{k}$ is within a spherical Fermi surface of radius $k_{\text{F}}$ due to the Pauli blocking effect.
We find that the binding energy $E_{\text{L}}$
is given by

\begin{eqnarray}
&&\frac{2}{\pi}\sqrt{-u_{\text{L}}}\arctan\left(\frac{\sqrt{-u_{\text{L}}}}{k_{\text{F}}}\right)+\frac{2k_{\text{F}}}{\pi} \nonumber \\
&=&\frac{1}{a_{\text{HL}}}+\frac{2}{\pi R} \int^{\infty}_{k_{\text{F}}} \! dq \ \frac{q \sin qR}{q^2-u_{\text{L}}}
\label{twobody}
\end{eqnarray}
where we have set $u_{\text{L}}=2m E_{\text{L}}/\hbar^2$.
This equation is valid for $-\infty<u_{\text{L}}<k^2_{\text{F}}$, and has
a unique solution for arbitrary $a_{\text{HL}}$.
One can easily verify that the solution can be written as
$u_{\text{L}}(a_{\text{HL}}, R, k_{\text{F}})=-k_{\text{F}}^2 g(k_{\text{F}} a_{\text{HL}}, k_{\text{F}} R)$,
where $g$ is a dimensionless scaling function.

In the absence of a Fermi sea or when $k_{\text{F}}\to0$, $u_{\text{L}}$ satisfies the following simple equation:
$\sqrt{-u_{\text{L}}}=\frac{1}{a_{\text{HL}}}+\frac{e^{-\sqrt{-u_{\text{L}}} R}}{R}$.
So for a positive scattering length and
$R=\infty$, $E_{\text{L}}=-\hbar^2 /2 m a^2_{\text{HL}}$ and
the light atom forms a two-body bound state with one of the heavy atoms.
However, when $R \ll |a_{\text{HL}}|$ the bound state is severely affected by the second
heavy atom and
$E_{\text{L}}=-\hbar^2 \Omega^2/2 m R^2$,
even when $a_{\text{HL}}$ is negative.
One can also show that right at resonance, as a result of non-linear interference
between waves coming off the two heavy atoms,
the effective scattering length for the light atom is proportional
to $R$, the distance between the two heavy atoms.

\begin{figure}
\begin{center}
\includegraphics[width=8.5cm]{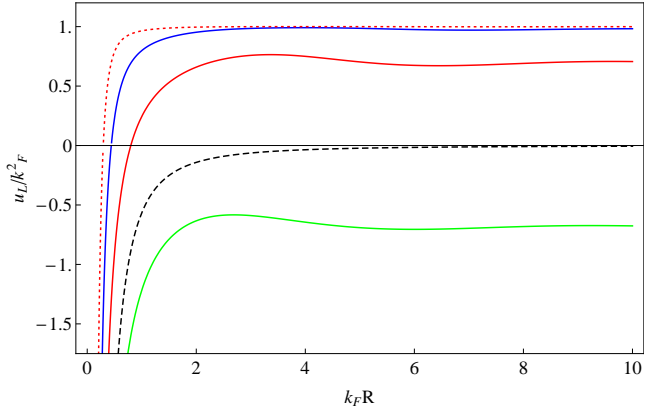}
    \caption{Effective potentials $u_{\text{L}}/k^2_{\text{F}}$ versus $k_{\text{F}}R$.
From top to bottom we plot the potential for $k_{\text{F}}a_{\text{HL}}=-1,\infty, 1$.
At resonance, $u_{\text{L}}(R)=-\Omega^2 /R^2$
if the Pauli blocking effect is absent (dashed line);
the dotted line is the estimated potential due to the scattering of trimers by the Fermi sea (see discussions on page 4).}
    \label{fig1}
\end{center}
\end{figure}

Now we turn to the effect of a finite density of fermions and focus on the
resonant case where $1/a_{\text{HL}}=0$; in this limit $g$ is a function of $k_{\text{F}} R$ only.
At short distances when $k_{\text{F}} R \ll 1$,
the effective attractive potential $u_{\text{L}}=-\Omega^2/R^2$ and is determined by the motion of a single light atom.
However, this universal behavior is completely
changed when $k_{\text{F}} R$ becomes order of unity or larger and the collective effect becomes important.
First at $k_{\text{F}} R=0.799$, the bound state energy $E_{\text{L}}$
increases to zero.
After this point, $E_{\text{L}}$
continues to rise until it eventually reaches a maximum and then decreases,
settling into a pattern of oscillations around a saturation value;
its asymptotic behavior is
\begin{eqnarray}
{u_{\text{L}}}(k_{\text{F}}R\to \infty) \to u^{\infty}_{\text{L}} \left(1-\frac{\cos k_{\text{F}}R}{(k_{\text{F}}R)^2}\right)
\end{eqnarray}
where $u^{\infty}_{\text{L}}\approx0.695 k^2_{\text{F}}$.
See Fig.\ref{fig1} where details are shown numerically.
Note that $\hbar^2 u^\infty_{\text{L}}/2\mu$ is
identical to the two-body binding energy at resonance and therefore represents
the atom-dimer threshold for trimers.

To understand the Pauli blocking effect on trimers,
we adopt a ${\bf k}$-space approach instead of employing the usual hyperspherical coordinates \cite{Efimov70}.
The Schr{\"o}dinger equation for a three-body wavefunction
$\Phi({\bf k}_1,{\bf k}_2,{\bf k}_3)$ is

\begin{eqnarray}
\left(\epsilon^{m}_{{\bf k}_1}+\epsilon^{M}_{{\bf k}_2}+\epsilon^{M}_{{\bf k}_3}-E\right)\Phi(\mathbf{k}_1,\mathbf{k}_2,\mathbf{k}_3)=
\frac{-V_0}{(2\pi)^3}\int d{\bf q}
\nonumber \\
\left( \Phi(\mathbf{k}_1-\mathbf{q},\mathbf{k}_2+\mathbf{q},\mathbf{k}_3)+
\Phi(\mathbf{k}_1-\mathbf{q},\mathbf{k}_2,\mathbf{k}_3+\mathbf{q})\right)
\end{eqnarray}
where $\mathbf{k}_1$ is the momentum of the light atom, $\mathbf{k}_2$ and $\mathbf{k}_3$ are the momenta for the heavy atoms and
$\epsilon^m_{{\bf k}_1}=\hbar^2{\bf k}^2_1/2m$, $\epsilon^M_{{\bf k}_2}=\hbar^2{\bf k}^2_2/2M$. $\Phi({\bf k}_1,{\bf k}_2,{\bf k}_3)$ vanishes when ${\bf k}_1$ is within the spherical Fermi surface of radius $k_{\text{F}}$. Furthermore, $\Phi({\bf k}_1, {\bf k}_2, {\bf k}_3)=\Phi({\bf k}_1, {\bf k}_3, {\bf k}_2)$ because of the Bose atom exchange statistics. For now we will consider the zero-total-momentum subspace
and further introduce a wavefunction for one heavy atom relative to the heavy-light dimer formed by the other two atoms:
\begin{displaymath}
\Psi_{\text{AD}}(\mathbf{k}_2)=\int_{|{\bf q}|>k_{\text{F}}} \! d\mathbf{q} \  \Phi({\bf q}, \mathbf{k}_2-{\bf q}, -\mathbf{k}_2).
\end{displaymath}
The resulting function $\Psi_{\text{AD}}(\mathbf{k})$ is shown to
satisfy

\begin{eqnarray}
\left(\frac{(2\pi)^3}{V_0}+
\int^{'} d\mathbf{q}
\frac{1}{\epsilon^{\mu}_{\mathbf{q}}+\epsilon^{\gamma}_{\bf k}-E}\right)\Psi_{\text{AD}}(\mathbf{k})\nonumber\\
=- \int^{'} d\mathbf{q} \frac{\Psi_{\text{AD}}(-{\bf q}-\frac{\mu}{m}\mathbf{k})}{\epsilon^{\mu}_{\mathbf{q}}+
\epsilon^{\gamma}_{{\bf k}}-E}
\end{eqnarray}
where $\gamma=M(M+m)/(2M+m)$ is the reduced mass for collisions between a heavy atom and a heavy-light dimer; also $\epsilon^{\gamma}_{\bf k}=\hbar^2{\bf k}^2/2\gamma$.
Integrals $\int^{'}$ are over
a region defined as $|-\frac{\mu}{M}\mathbf{k}+\mathbf{q}| > k_{\text{F}}$
to exclude the occupied states.
When $\frac{\mu}{m}$ approaches $1$, or in the Born-Oppenheimer approximation,
we can integrate out the fast degrees of the light atom and map the three-body problem to a simple equation for
the dimer and heavy atom:

\begin{eqnarray}
&&\left(\frac{2}{\pi}\sqrt{\frac{k^2}{\beta^2}-u_{\text{B}}}\arctan\left(\frac{\sqrt{\frac{k^2}{\beta^2}-u_{\text{B}}}}{k_{\text{F}}}\right)+\frac{2k_{\text{F}}}{\pi}-\frac{1}{a_{\text{HL}}}\right)
\Psi_{\text{AD}}(\mathbf{k})
\nonumber \\
&&=\frac{1}{2\pi^2}\int\limits_{q>k_{\text{F}}} \! d\mathbf{q}\ \frac{\Psi_{\text{AD}}(\mathbf{q}-\mathbf{k})}{q^2+k^2/\beta^2-u_{\text{B}}}.
\label{threebody}
\end{eqnarray}
Eq.\ref{threebody} is valid for $-\infty<u_{\text{B}}<k^2_{\text{F}}$.
Here $u_{\text{B}}=2\mu E/\hbar^2$ and $\beta=\sqrt{\gamma/\mu}$ which is much larger than one in our case.
The kinetic term (the square-root term on the left hand side of the equation)
scales linearly as a function of $k$ for $k \gg \beta \sqrt{u_{\text{B}}}$,
reflecting the composite nature of the dimer at short distance;
this peculiar structure was appreciated in earlier studies on dimer-atom scattering \cite{Skorniakov57,Petrov04}.
One can also write down the corresponding differential equation for the Fourier transformed wavefunction
$\Psi_{\text{AD}}({\bf r})=\int d{\bf k}\Psi_{\text{AD}}({\bf k})\exp(i{\bf k}\cdot {\bf r})$.
In the limit where $\beta$ is infinite, Eq.\ref{threebody} for $\Psi_{\text{AD}}({\bf r}={\bf R})$ is equivalent to Eq.3 for two heavy atoms at a fixed distance $|{\bf R}|=R$ apart.

To obtain the spectrum flow of Efimov states,
we carry out a WKB analysis of the wavefunction $\Psi_{\text{AD}}({\bf r})=\exp(i\beta S_0(r)+iS_1(r))$ and 
the bound state energies $u_{\text{B}}$; the WKB approach is valid
as far as
$\beta$ ($\sim \! \! \sqrt{M/2m}$) is much larger than unity.
By comparing the resulting equation for $S_0$, derived 
from Eq.\ref{threebody}, to 
Eq.\ref{twobody}, one further
establishes that $S_0$ is a simple function of $u_{\text{L}}(r)$.
For the $S$-wave channel, $\Psi_{\text{AD}}(r)=\frac{e^{i S_1(r)}}{r} \sin\left(\beta\int^{r_1}_{r} \! dr'\sqrt{u_{\text{B}}-u_{\text{L}}(r)}+\delta\right)$
for $r< r_1$, where $r_1$ is the first semiclassical turning point defined by $u_{\text{L}}(r_1)=u_{\text{B}}$  \cite{S1}
and $\delta$ is a phase shift calculated below.
When $r$ approaches zero,
the phase integral has a logarithmic divergence for any value of $k_{\text{F}}$, as seen in the hyperspherical approach to the three-atom problem.
Without losing generality,
we proceed by introducing a Bethe-Peierls boundary condition
at $r=r_0$
to take into account a hard-core repulsion with range $r_0$
that is much shorter than the Fermi wavelength $2\pi/k_{\text{F}}$.
The quantization condition is:
\begin{eqnarray}
 w(k_{\text{F}}r_0, \frac{u_{\text{n}}}{k_{\text{F}}^2})=\int^{r_1}_{r_0} \! dr'\sqrt{u_{\text{n}}-u_{\text{L}}(r')}=\frac{1}{\beta}(n\pi-\delta)
\end{eqnarray}
where $n=1,2,3,...$ indexes the eigenvalues.
Since $u_{\text{L}}(r)/k_{\text{F}}^2$ is only a function of $k_{\text{F}} r$, the dimensionles phase $w$ is only a function of $k_{\text{F}}r_0$ and $u_{\text{B}}/k_{\text{F}}^2$.
$\delta$ can be obtained by
applying the standard matching formulas for the Schr{\"o}dinger equation, and if there is a single turning point we get $\delta=\pi/4$. \cite{phase}.

\begin{figure}
\begin{center}
        \includegraphics[width=8.5cm]{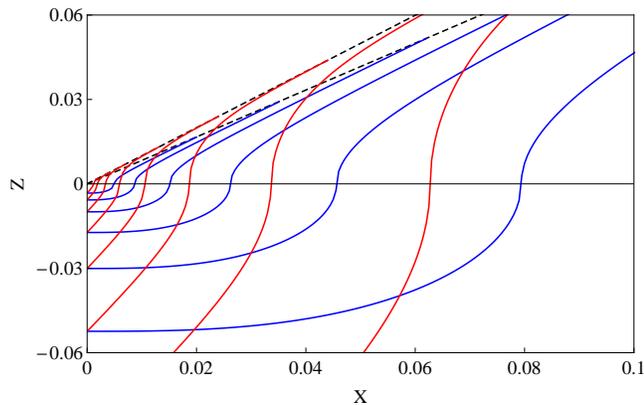}
    \caption{(color online) A subset of eigenvalue trajectories at resonance ($a_{\text{HL}}=\infty$) in the $Z=
\text{sgn}(u_{\text{B}}) \sqrt{|u_{\text{B}}|}r_0$ versus $X=k_{\text{F}}r_0$ plane;
the red trajectories are for trimers dressed in fluctuating particle-hole pairs estimated in the leading order. The ray $Z=\sqrt{Y^{\infty}} X$ (dashed line)
represents the dimer-atom threshold. The termination points along this threshold $X_n^{\infty}$ are discussed in Eq.\ref{termination}. Here $\beta=\sqrt{\gamma/\mu}=10$, and $r_0$ is the hard-core size of Bose atoms. Only Efimov states are shown, and we take $\delta=\pi/4$ \cite{phase}.}
    \label{fig3}
\end{center}
\end{figure}

The simplest case is where $k_{\text{F}}=0$.
Then there are infinitely many solutions that accumulate near zero energy, and
$u_{n}/u_{n+1}\to e^{\frac{2\pi}{\Omega \beta}}$ as $n\to\infty$, with
$e^{\frac{\pi}{\Omega}}\approx 254.5$ \cite{Efimov70,Efimov72}. For a non-zero fermion density, we introduce three dimensionless parameters:
$X=k_{\text{F}}r_0$, $Y=u_{\text{B}}/k^2_{\text{F}}$ and $Z=\text{sign}(Y) \sqrt{|Y|}X$. Our results on the spectrum flow are presented in Fig.\ref{fig3} in the
$Z$-$X$ plane, where each eigenvalue trajectory shows how the solution to Eq.8 for fixed $n$ 
varies as we increase $X$.  
For fixed non-zero $X$, the ratio $u_n/u_{n+1}$ depends on $n$ and 
approaches the universal value of $\exp(2\pi/\Omega\beta)$ only for 
low lying states with $n$ much less than $N(X)$, 
where $N(X)$ denotes the number of Efimov states at a given density $X$. 
Since $N(X)$ is a rapidly decreasing function of $X$, 
the universal value can only be attained when $X$ is small. 
As $X$ increases, each eigenvalue increases towards $u^{\infty}_{\text{L}}$, the dimer-atom threshold, eventually reaching this value and disappearing from the spectrum; the eigenvalue trajectories terminate along the ray $Z=X \sqrt{u^{\infty}_{\text{L}}/k^2_{\text{F}}}$ in the $Z$-$X$ plane.  The termination point $X^{\infty}_n$ at which $u_n$ reaches the dimer-atom threshold is given by the solution to:
\begin{eqnarray}
w(X^{\infty}_n, Y^\infty)=\frac{1}{\beta}\left(n\pi-\delta^{\infty} \right)
\end{eqnarray}
where $Y^\infty=u^{\infty}_{\text{L}}/k^2_{\text{F}}=0.695$ and $\delta^\infty$ is the phase shift at $Y=Y^\infty$.

As far as $X$ is much less than unity, the
termination points have certain universal properties.
Indeed, as $X$ approaches $0$, $w(X,Y)$ approaches an asymptotic form $w(X,Y)\to -\Omega \log X + v(Y)$
where $v(Y)$ is a regular function of $Y$ for $-\infty<Y<Y^{\infty}$.
This formula yields estimates for several quantities of interest.
For example, from Eq.8 we find that 
%$N(X)=\lfloor\frac{\beta}{\pi}w(X,Y^{\infty})+\frac{\delta}{\pi}\rfloor$, 
$N(X)\to\frac{\beta\Omega}{\pi}\log\frac{e^{\frac{v(Y^{\infty})}{\Omega}}}{X}+\frac{\delta^{\infty}}{\pi}$ as $X$ approaches $0$. 
Furthermore, the termination points are
$X^{\infty}_n\approx e^{\frac{1}{\Omega}\left(v(Y^{\infty})+\frac{\delta^{\infty}}{\beta}\right)}e^{-\frac{n \pi}{\Omega \beta}}$
for small $X^{\infty}_n$, which leads to a universal relation between different termination points:

\begin{eqnarray}
X^{\infty}_n/X^{\infty}_{n+1}= e^{\frac{\pi}{\Omega \beta}},
\label{termination}
\end{eqnarray}
for small $X^{\infty}_n$ or large $n$, and otherwise independent of $n$.  We can also calculate $X_n(Y)$, the point at which $u_n$ increases to $Y$, for $-\infty<Y<Y^{\infty}$; we obtain a result identical to Eq.10 after replacing $X^{\infty}_n/X^{\infty}_{n+1}$ with $X_n(Y)/X_{n+1}(Y)$ in that expression. These are the new universal properties in three-body physics
when there is a Fermi sea quantum background. Our numerical results agree well with the WKB results: for $\beta$ = 10, the difference is within a few percent for the densities shown in Fig. \ref{fig3}.

These universal structures in the
spectrum flow should be robust against the scattering of trimers by the Fermi sea.
So far we have treated the Fermi sea as static or incompressible,
but a trimer can further collide with the Fermi sea, exciting 
particle-hole pairs near the Fermi surface and polarizing the background.  
One of the main effects of these fluctuating pairs is to 
suppress the interspecies interactions, as occurs in an interacting Fermi gas \cite{Gorkov61}, leading to a reduction of the Fermi-Bose dimer binding energy (estimated in Ref.\cite{Song10}). 
Following those diagrammatic calculations, in the limit of heavy Bose atoms
one finds that the dimer binding energy measured from the Fermi energy is reduced
by a factor of $m/M$ when $k_{\text{F}} a_{\text{HL}}$ is small and negative. This can also be attributed to an effect related to Anderson's infrared
catastrophe \cite{Anderson67}. As a result, $Y^\infty$, 
which specfies the dimer-atom threshold for trimers, moves closer to unity (see Fig.\ref{fig1}). $Y^\infty$ and $v(Y^\infty)$ are now functions of $\beta$, and 
these mass-ratio-dependent corrections to the effective potential shift the flow of each eigenvalue trajectory. However, they do not
affect the universal relation between different trajectories, as Eq.\ref{termination} is independent of $Y^\infty$ and $v(Y^\infty)$; we obtain the same universal value for $X^{\infty}_n/X^{\infty}_{n+1}$, provided $X^{\infty}_n$ is small. 
At densities close to $X_n^{\infty}$ and related by Eq.\ref{termination}, 
we anticipate distinct peak-dip signatures in the inelastic loss spectrum similar to those observed in Ref.\cite{Kraemer06,Ottenstein08,Williams09,Ferlaino09,Zaccanti09,Pollack09}.

In conclusion, we have shown that the spectrum flow of Efimov states when the density of fermions varies is universal
as far as the hard-core size of heavy Bose atoms $r_0$ is much shorter than
the average interatomic distance. Our results can be applied to understand three-body states in
$^{87}\text{Rb}$-$^{6}\text{Li}$ and $^{23}\text{Na}$-$^{6}\text{Li}$ mixtures.
It is possible to generalize the above idea to other situations.
For instance, when applying the same approach to the case of one light boson in resonance with
two heavy Fermi atoms in the presence of a Fermi sea,
we find that
because of the anti-screening effect of background Fermi atoms,
the mediated potential $u_L(r)=-\Omega^2/r^2 [1-q_0r si(q_0r)]$,
where $q_0=\beta^2 k_{\text{F}}/(2\pi)^2$ and $si(x)$ is a sine integral which oscillates and approaches $-\frac{\cos x}{x}(1-\frac{2}{x^2})$ when $x \to \infty$.
A similar spectrum flow can be obtained and details will be presented elsewhere.
There are also open questions that need to be answered in the future.
One is the role of hole-like configurations and the effect of Fermi sea polarization
on the three-body spectrum \cite{Nishida09}.
Another question is what are the dynamic consequences of these Efimov states embedded in a quantum gas and
how do they affect the inelastic loss spectrum and dimer-atom or dimer-dimer elastic
scattering \cite{Petrov04}?
The eventual fate of such a quantum gas after a large fraction of
scattering atoms fall into
the Efimov channel remains unknown.
A straightforward way to proceed
is perhaps to include more atoms in the Efimov channel
and extend the analysis to 4-body, 5-body etc \cite{Hammer07,Stecher09}.
Along that road, quantum
Monte Carlo simulations have been performed to
understand cluster structures
of bosonic atoms \cite{Hanna06,Stecher10}.
Similar simulations need to be performed in the presence
of background scattering atoms.
This work is supported by NSERC (Canada) and Canadian Institute for Advanced Research.
We thank Eric Braaten, Aurel Bulgac,
Chris Greene, Yusuke Nishida, Junliang Song and Peter Zoller for discussions on Efimov states.


\begin{references}{}



\bibitem{Efimov70}V. Efimov, Phys. Lett. B{\bf 33}, 563(1970);
Sov. Jour. Nucl. Phys. {\bf 12}, 589(1971).
\bibitem{Efimov72}
V. Efimov, JETP Lett.{\bf 16}, 34 (1972).

\bibitem{Esry99}B. D. Esry {\it et. al}, Phys. Rev. Lett {\bf 83}, 1751(1999).
\bibitem{Braaten01}E. Braaten {\it et al.}, Phys. Rev. Lett.{\it 87}, 160407 (2001).
\bibitem{Braaten06}P. F. Bedaque {\it et al.}, Phys. Rev. Lett. {\bf 82}, 463(1999).
\bibitem{Hammer07}H. W. Hammer, L. Platter, Eur. Phys. J. A32, 113 (2007).

\bibitem{Stecher09}
J. Von Stecher {\it et al.}, Nat. Phys. {\bf 5}, 417 (2009).
%J. Von Stecher, J. P. D'Incao, C. H. Greene, Nat. Phys. {\bf 5}, 417 (2009).

\bibitem{Kraemer06}T. Kraemer et al., Nature {\bf 440}, 315 (2006).
\bibitem{Ottenstein08}T. B. Ottenstein, Phys. Rev. Lett. {\bf 101}, 203202(2008).
\bibitem{Williams09}J. R. Williams {\it et al.}, Phys. Rev. Lett. {\bf 103}, 130404(2009).
\bibitem{Ferlaino09}F. Ferlaino {\it et al.}, Phys. Rev. Lett. {\bf 102}, 140401 (2009).
\bibitem{Zaccanti09}M. Zaccanti {\it et al.}, Nat. Phys. {\bf 5}, 586 (2009).
\bibitem{Pollack09}S. E. Pollack {\it et al.}, Science {\bf 326}, 1683 (2009).
%\bibitem{Pollack09}S. E. Pollack, D. Dries and R. G. Hulet, Science {\bf 326}, 1683 (2009).

%\bibitem{Greene10}Chris Greene, Physics Today {\bf 63}, 5 (2010).

\bibitem{Skorniakov57}G. V. Skorniakov and K. A. Ter-Martirosian, Sov. Phys.
JETP {\bf 4}, 648 (1957).
\bibitem{Faddeev61}L. P. Faddeev, Sov. Phys. JETP {\bf 12},1014(1961).
\bibitem{Cui10}$a_{HL}$ sets the boundary condition of the RG flow; see X. L. Cui {\it et al.}, Phys. Rev. Lett.{\bf 104}, 153201 (2010).
\bibitem{Petrov04}
D. S. Petrov {\it et al.}, Phys. Rev. Lett. {\bf 93},
090404 (2004).
%D. S. Petrov, C. Salomon and G. V. Shlyapnikov, Phys. Rev. Lett. {\bf 93},090404 (2004).




\bibitem{S1}
The real function $iS_1$ represents the next order
contribution (in terms of $1/\beta$) that we don't show explicitly. 
When $r$ approaches zero, $\exp{iS_1(r)}\sim |u_{\text{L}}(r)|^{(1+\Omega-\Omega^2)/4(1-\Omega)} /|u_{\text{B}}(r)-u_{\text{L}}(r)|^{1/4}$.

\bibitem{phase}
Since $u_{\text{L}}(r)$ oscillates around its asymptotic value,
there can be multiple turning points
when $u_n$ is close to $u^\infty_L$ and
$\delta$ can depend on $u_{n}$. For instance,
if there are additional turning points at $r_2$ and $r_3$,
the phase shift is
$\delta=\frac{\pi}{4}+\arctan\left(\frac{1}{4}\tan\theta e^{-2\eta}\right)$
with $\theta=\beta\int^{r_3}_{r_2} \! dr'\sqrt{u_n-u_{\text{L}}(r')}$ and $\eta=\beta\int^{r_2}_{r_1}
\! dr'|\sqrt{u_{n}-u_{\text{L}}(r')}|$.
For typical values of $u_n$ (i.e. $\theta$ is away from $\pi/2$) and large $\beta$,
the deviation of $\delta$ from $\pi/4$ is exponentially small.
However, for specific values of $u_n$,
$\tan\theta$ can be large indicating additional non-Efimov states at these energies,
bound in the oscillations at large $k_F r$.
Very near these values, 
$\delta$
has a strong $u_n$ dependence 
and Eq.8 can have multiple solutions for a given
$n$ reflecting the emergence of non-Efimov bound states (not shown in Fig.\ref{fig3}).
However, the hybridization between these and the Efimov states shown in Fig.\ref{fig3} is exponentially suppressed in the large $\beta$ limit.
In this Letter, we only consider states in the Efimov family (which arise from the $k_{\text{F}}\to0$ bound states). These complications do not affect the universal structures discussed below.
\bibitem{Gorkov61}L.P. Gorkov and T. K. Melik-Barkudarov, Sov. Phys. JETP {\bf 13}, 1018(1961).
\bibitem{Song10}Junliang Song {\it et al.},
Phys. Rev. Lett. {\bf 105}, 195301 (2010).
\bibitem{Anderson67}
P. W. Anderson, Phys. Rev. Lett.{\bf 18}, 1049(1967).

\bibitem{Nishida09}
After submission, we learned that 
the polarization effect has been discussed in the context of an impurity problem,
Yusuke Nishida, Phys. Rev. A {\bf 79}, 013629(2009).


\bibitem{Hanna06}G. J. Hanna, D. Blume, Phys. Rev. A{\bf 74}, 063604 (2006).
\bibitem{Stecher10} J. Von Stecher, Jour. Phys. B: {\it At. Mol. Opt. Phys.} {\bf 43}, 101002 (2010).



\end{references}
\end{document}